\documentclass[printer]{aa}

\newcommand{\peryr}{\,{\rm yr^{-1}}}
\newcommand{\pers}{\,{\rm s^{-1}}}
\newcommand{\erg}{\,{\rm erg}}

\newcommand{\Gyr}{{\,\rm Gyr}}

\newcommand{\Mo}{M_{\odot}}

\def\gsim{ \lower .75ex \hbox{$\sim$} \llap{\raise .27ex \hbox{$>$}} }
\def\lsim{ \lower .75ex\hbox{$\sim$} \llap{\raise .27ex \hbox{$<$}} }

\usepackage{graphicx}
\authorrunning{Guetta \& Stella}
\titlerunning{GRAVITATIONAL WAVES FROM SHORT GRBs}

\begin{document}

%\shorttitle{GRAVITATIONAL WAVES FROM SHORT GRBs}
%\shortauthors{Guetta and Stella}

\title{Short $\gamma$-ray Bursts and Gravitational Waves from Dynamically Formed 
Merging Binaries}

\author{Dafne Guetta\inst{1} and Luigi Stella\inst{1}}

\institute{Osservatorio astronomico di Roma, v. Frascati 33,
00040 Monte Porzio Catone, Italy.}

\abstract{
Merging binary systems consisting of two collapsed objects 
are among the most promising sources of high frequency gravitational 
wave, GW, signals for ground based interferometers. 
Double neutron star or black hole/neutron star mergers are also believed 
to give rise to short hard bursts, SHBs, a subclass of gamma ray bursts.
SHBs might thus provide a powerful way to 
infer the merger rate of two-collapsed object binaries.
Under the hypothesis that most SHBs originate from 
double neutron star or black hole/neutron star mergers, 
we outline here a method to estimate the incidence of merging 
events from dynamically formed binaries in globular clusters and 
infer the corresponding GW event rate that can be detected with 
Advanced LIGO/Virgo. In particular a sizeable fraction of 
detectable GW events is expected to be coincident with SHBs. 
The beaming and redshift distribution of SHBs are reassessed 
and their luminosity function constrained by using the results 
from recent SHBs observations.  We confirm that a substantial 
fraction of SHBs goes off at low redshifts,
where the merging of systems formed in globular clusters through dynamical 
interactions is expected. 

}
\maketitle

\section{Introduction}

Merging binary systems containing two collapsed objects, {\it i.e.} a double neutron star (DNS),  
a stellar-mass black hole and neutron star (BH-NS) or two stellar-mass black hole (BH-BH),
are powerful gravitational wave, GW, sources with frequencies from hundreds 
to over a thousand Hz. They are considered 
among the most promising GW sources for ground-based interferometers,
of the current and future generation, 
such as LIGO, Virgo and their advanced versions. 
BH-NS and, especially, BH-BH mergers 
emit more powerful and lower frequency GWs  
than DNS mergers, where the sensitivity of 
LIGO and Virgo detectors is highest: therefore they 
can be detected up to larger distances. 
The horizon of first generation LIGO and Virgo for DNSs, BH-NS and BH-BH  
mergers is $\sim 20$, $43$ and $100$ Mpc, respectively, 
while Advanced LIGO/Virgo class interferometers 
should detect them out to a distance 
of $\sim 300$, $650$ and $1600$ Mpc 
(for a review see Cutler \& Thorne 2002). 

The rate of detectable merging events
has been estimated based on the observed galactic
population of DNS binaries containing a radio pulsar (Phinney 1991; 
Narayan, Piran \& Shemi 1991; Kalogera et al. 2001; Burgay et al. 2003).
The best estimate of the DNS merger rate 
in the Galaxy is presently 
$\sim 80^{+200}_{-60}$/Myr, converting to 
$800^{+2000}_{-600}$ Gpc$^{-3}$yr$^{-1}$ 
for a galaxy number density of $10^{-2}$Mpc$^{-3}$ (Kalogera et al. 2004). 
Population synthesis studies of binary systems give results consistent
with the above rate (Perna \& Belczynski 2002; Belczynski, Kalogera \& Bulik 2002; 
Belczynski et al. 2007).  
GW signals from DNS mergers are expected at a rate of one in $\sim$ 10-150
years with Virgo and LIGO and one every 1-15 days with Advanced LIGO/Virgo 
class interferometers.
The BH-NS and BH-BH merger rates in the Galaxy
are highly uncertain. Belczynski et al. (2007)
estimate $\sim 1\%$ and $\sim 0.1\%$ of the DNS merger rate, respectively,
implying that BH-NS and BH-BH mergers contribute marginally to the 
GW event rate, despite the 
larger distance up to which they can be detected.
 
DNSs and NS-BH merging events provide also one of the leading models 
for Short Hard Gamma Ray Bursts, SHBs, bursts of $\gamma$-rays that last 
$< 2$~s and take place at cosmological distances (Goodman 1986; Paczy\'nski 1986; Eichler et
al. 1989; Narayan, Paczynski \& Piran 1992).  As with the collapsar scenario 
for long Gamma Ray Bursts, GRBs (Woosley \& Mac Fadyen  1999), such models
envisage the formation of a black hole surrounded by a torus of matter at nuclear density
that is rapidly accreted and provides the primary source of energy for the burst.
For this reason it is believed that BH-BH mergers are not among the progenitors
of SHBs. 
SHBs comprise about 1/4 and 1/10 of the CGRO/BATSE and {\it Swift}/BAT GRB samples.
The X-ray and optical afterglows of SHBs have been observed only recently, 
leading to the first
identifications and redshift determinations of SHB host galaxies 
(see e.g.  Berger et al. 2007a and references therein).
Out of the $\sim 30$ SHBs in the current {\it Swift} 
and HETE~2 sample
about half displayed an optical afterglow, $11$ have redshift, 
and a few show evidence for beaming in their optical afterglow light curve
(Fox et al. 2005; Soderberg et al. 2006).
The association of some SHBs with galaxies characterized by a very low star 
formation rate (SFR, of order $\lsim 0.1\Mo\peryr$) indicates  a long time-delay 
between the progenitors' collapse and time at which these SHBs go off. 

It has long been known that DNSs and BH-NS systems (hereafter NS-NS/BH systems), 
can form from massive binaries surviving two gravitational collapses
(we term these ``primordial'' NS-NS/BH systems).
Population synthesis calculations show that 
primordial NS-NS/BH systems merge after a relatively short 
time ($\tau\!\approx\!2\Gyr$ on average), implying that the 
redshift-distribution of SHBs should follow pretty closely the 
SFR history of the universe (e.g. Bloom, Sigurdsson \& Pols 1999; 
Belczynski et al. 2002, 2006). 
Alternatively NS-NS/BH systems can form through
dynamical interactions in the cores of globular clusters, GCs. 
The dominant process comprises two stages: 
first an isolated NS (or BH) in the cluster captures a 
non-degenerate star during a close encounter, thus forming a binary
containing a collapsed object. 
This binary then undergoes an exchange
interaction with another collapsed object, leading to the ejection of the 
non-degenerate star and formation of a NS-NS/BH binary  
(Verbunt \& Hut 1987; Efremov  2000). 
Grindlay et al. (2006) estimate that $\sim 10-30\%$ of all DNS mergers 
may stem from such dynamically formed systems. 
Dynamical interactions take place efficiently at high stellar densities, 
which, in turn, are highest in the core of globular cluster that have 
undergone core collapse. Therefore, the delay between the formation 
of the collapsed objects and merging of dynamically-formed NS-NS/BH systems  
is dominated by the time until core-collapse, which is
typically comparable to the Hubble time.
In their simulations Sadowski et al. (2008) found 
that merging BH-BH binaries form rather efficiently
in dense clusters, but failed to find any merging of NS-NS/BH binaries.

If NS-NS/BH mergers are among the progenitors of SHBs,  
the SHB rate can be used as an alternative way of constraining the 
DNSs merger rate (Guetta and Piran 2005, 2006 (GP05, GP06), Nakar et al. 2006). 
Hopman et al (2006)(H06) showed that dynamically formed DNSs give rise to more 
numerous low-z bursts than would be expected if SHB closely followed the SFR,  
as in the case of primordial DNSs. The z-distribution of dynamically 
formed DNSs provides a better match to observed redshifts of SHBs.
This result was recently confirmed 
by Salvaterra et al. (2008), who  also conclude that SHBs may well originate 
from both classes of DNSs.
GP05, GP06 and H06 fitted the peak flux distribution of 
SHBs detected by BATSE to derive 
their formation rate and luminosity function (LF), 
in both the dynamically-formed and primordial DNS scenarios, while 
Nakar et al. (2006) and GP06 studied the 
prospects for detecting GW signals from SHBs from primordial DNSs. 

In this paper, based on the results from recent SHB observations,  
we first reassess the beaming, redshift distribution and LF of SHBs. 
Under the hypothesis NS-NS/BH mergers give rise to most SHBs, 
we then present a method to estimate the incidence of merging 
events from dynamically formed binaries in globular clusters and 
infer the corresponding GW event rate that can be detected with 
Advanced LIGO/Virgo class interferometers. 
 
\section{Luminosity function and rate evolution of SHBs}\label{LF}

For primordial NS-NS/BH systems, the intrinsic SHB rate is given by the 
convolution of the formation rate of NS-NS/BH binaries
(which is assumed to follow with negligible delay the SFR) 
with the distribution of the merging time delays, $P_{\rm pr}(\tau)$, 
where $\tau = \tau_{\rm GW}$ is the time over which GW losses bring a 
binary to its pre-merging stage.  
We adopt $P_{\rm pr}(\tau)\sim 1/\tau$
in agreement with 
the $\tau$-values of DNS radio pulsar binaries and population 
synthesis calculations (Champion et al. 2004; Belczynski et al. 2007).
 
The SHB rate from primordial DNSs is given by
\begin{equation}
\label{ratep} R_{\rm SHB}(z) \propto  \int_{0}^{t(z)} d\tau R_{\rm
SF2}(t-\tau) P_{\rm pr}(\tau) ,
\end{equation}
where $R_{\rm SF2} $ refers to the ``SF2" model for the SFR in Porciani \& Madau (2001).

For dynamically formed NS-NS/BH binary, $\tau = \tau_{\rm CC} + \tau_{\rm GW}$, 
where the delay time $\tau_{\rm CC} \gg \tau_{\rm GW}$ represents the  
elapsed time between the birth of NSs and BHs in GCs and 
the dynamical formation of NS-NS/BH systems following core collapse.  
According to H06, the distribution of dynamically formed 
NS-NS/BH binaries $P_{\rm dyn}(\tau)$ (which replaces $P_{\rm pr}(\tau)$
in Eq.1) 
increases for increasing time delays.
We assume that the formation rate of GCs is proportional 
to the total SFR \footnote{
We note that Salvaterra et al. 2008 considered two alternative GC 
formation rates and found that these produce very similar results to 
those obtained by assuming that GC formation follows the SFR.}.
We derive an average delay of $\bar{\tau}\approx 2$ and $\approx6\Gyr$,
for primordial and dynamically formed 
NS-NS/BH system, respectively (for the latter we considered only 
values of $\tau$ shorter than the Hubble time). 

\subsection{The SHB luminosity function}

In this section we summarize the method adopted by GP06 and H06
to estimate the SHBs' LF and we derive the local SHB rates 
independently for primordial and dynamically formed NS-NS/BH binaries.
We use here the same sample of 194 SHBs detected by BATSE as in 
Paciesas et al. (1999).
GP06 and H06 modelled the ``isotropic-equivalent" LF ({\it i.e.}  
uncorrected for beaming), with a broken power law extending 
from $L^*/\Delta_1$ to $L^*\Delta_2$ and with break at $L^*$
(with $\Delta_{1}=\Delta_{2}=100$), i.e.
\begin{equation}
\label{Lfun} \Phi_o(L) d\log L =C_0 d\log L \left\{
\begin{array}{ll}
(L/L^*)^{-\alpha}; & L^*/\Delta_1 \!<\! L \!<\! L^* \\ (L/L^*)^{-\beta}; & L^* \!<\!
L \!<\! \Delta_2 L^* 
\end{array}
\right. \ 
\end{equation}
($C_0$ is a normalization constant).
The best fit LF was derived by using the SHB rate $R_{\rm SHB}(z) $
for the two formation scenarios (Eq. 1).
Using these ``best fit" LFs these authors calculated 
the predicted redshift distribution in the two cases and 
compared it with present sample of SHB redshifts (see Sect 2.2). 
By using the same procedure, we rederived 
the best fit parameters for the LF and determined for the 
first time the local event rate, $R_0$ (see Table 1), 
independently in the two scenarios. 
We find that, by virtue of their longer $\tau$, dynamically 
formed systems would have to merge at a $\sim 3$ times higher rate 
than primordial systems in the local universe. 

The results in Table~1 are virtually insensitive to an increase of 
$\Delta_2$, whereas $L_{\rm min}=L^*/\Delta_1$ cannot be increased by more 
than a factor of 10, 
without excluding the least luminous SHB detected so far 
(GRB050509B, Gehrels et al. 2005).
By decreasing $L_{\rm min}=L^*/\Delta_1$, 
the local SHB rate increases approximately as  
$R_0\propto (L_{\rm min})^{\alpha}$ (Nakar 2006).
Objects with luminosity $L_{\rm min}$ observed by CGRO/BATSE or 
{\it Swift}/BAT at a limiting flux of 
$F_{\rm lim}$,  are detectable up to a maximum redshift of 
$z_{\rm max}$ which decreases with decreasing $L_{\rm min}$
(for {\it Swift}/BAT 
$F_{\rm lim}$ is $\sim 1\,{\rm ph\, cm^{-2}\,s^{-1}}$; Sakamoto et al. 2008).
Therefore SHBs with very low
luminosity are above the limiting sensitivity of BATSE and BAT
only in a fairly small volume, where they go off rarely.
Current data do not constrain this part of the LF and
the number of very weak SHBs (if any) remains unknown. 
It cannot be excluded that the volume afforded by the limiting sensitivity of 
BATSE and BAT is close to that required in order to sample 
the low-luminosity end of the LF of SHBs over the satellites' lifetime.
However this appears contrived and the low-luminosity
end of the SHB LF might well extend below the BATSE and BAT sensitivity. 
We note in passing that the presently-known LF of long GRBs extends 
over more than 4 decades
(see e.g.  Liang et al. 2007, Guetta \& Della Valle 2007). 
In Sect 3 we discuss further the implications 
of a SBH LF that extends below the currently-estimated value.

There exists by now evidence that SHBs 
are beamed in a relatively small solid angle. 
Similar to the case of long GRBs, 
the jet opening angle is inferred from the steepening of the optical 
afterglow light curve. Fox et al. (2005) found
a beaming factor of $f_b^{-1}\sim 50$ for GRB 050709 and GRB 050724 
(with $f_b$ the fraction of the $4\pi$ solid angle within which the GRB 
is emitted).
Soderberg et al (2006) inferred $f_b^{-1}\sim 130$ for GRB 051221A.
We adopt a fiducial value of $f_b^{-1}\sim 100$ and derive a beaming-corrected rate of  
$\rho_0 =f_b^{-1} R_0\sim 130(f_b^{-1}/100)$ and  $\sim 400 (f_b^{-1}/100)$ 
Gpc$^{-3} $yr$^{-1}$ for 
SHBs originating in primordial and dynamically formed binaries, respectively.
The former estimate compares well with the lower end of the range estimated by 
Belczynski et al. (2002) and 
Kalogera et al. (2004) for primordial DNSs ( $200 - 2800 $Gpc$^{-3}$yr$^{-1}$). 
 
\subsection{The observed $z$ distribution of SHBs}

In order to infer the redshift distribution of SHBs, we use the 
entire sample of presently known SHBs with redshift determination, 
{\it i.e.} the sample of events in Table~2 of 
Salvaterra et al. (2008) 
plus GRB~071227, GRB~070429B and GRB~070714 at z = 0.384, z=0.902 and z=0.922, 
respectively (D'Avanzo et al. 2007;  Berger et al. 2007b; 
D'Avanzo et al. 2008;  Cenko et al. 2008). 
We excluded two of the bursts (GRB~061006, GRB~061201) from the sample 
of Salvaterra et al. (2008) because uncertainties in the redshift determination  
\footnote{However we note that D'Avanzo et al. (2008) have recently
confirmed the redshift of GRB~061201.} 
We have verified that re-inserting these bursts in our sample 
does not affect significantly any of our results.

In Figure 1 we compare the cumulative distribution of the 
observed redshifts with the expected cumulative 
$z-$distribution given by 
\begin{equation}
\label{redshift} N_{\rm exp}(z)= \frac{R_{\rm SHB}(z)}{1+z} \frac{dV(z)}{dz}
\int_{L_m}^{L_{\rm max}} \Phi_o(L)d\log L \
 ,
\end{equation}
where $L_{\rm max}=\Delta_2 L^* = 100L^*$ and $L_m={\rm max}(L_{\rm F}, L^*/\Delta_1)$, 
with $L_{\rm F}$ the luminosity at $z$ corresponding to $F_{\rm lim}$.
This is done in turn for the primordial and dynamical
formation scenarios, by using the parameters given in Table 1 
and $\Delta_1 = 100$ (see Fig.~1).
A Kolmogorov-Smirnov (KS) test gives a probability of 0.04 and 0.2 
that the observed distribution is drawn from the distribution 
expected for SHBs from primordial and dynamically formed 
NS-NS/BH binaries, respectively. 
These results are nearly insensitive 
to increasing values of $\Delta_1$ (see also in Sect.~2.1).

We then considered the cumulative z-distribution that results
from a combination of primordial and dynamically,
formed NS-NS/BH systems. By varying their relative incidence 
the closest matching with the observed z-distribution
of SHBs was found for a $\sim 60$\% contribution from dynamically
formed NS-NS/BH binaries (see the dot-dashed line in Figure~1). 
In order to get a rough 
lower limit on the incidence of dynamically formed NS-NS/BH systems
we decreased their contribution until a KS probability of 0.1 was 
reached: this gives a $\sim 10$\% incidence of such systems. 
 We note that the corresponding local SHB rate is $R_0 \sim 2.9$
and $1.6$~Gpc$^{-3}$yr$^{-1}$ for a 60\% and 10\% fraction of 
dynamically formed mergers, respectively.

Of course this incidence can be derived
only approximately, in consideration of the small sample of SHBs 
with redshift determination known so far. 
We emphasise that the relative contribution of the 
two populations of merging binaries is determined 
primarily by the observed redshift distribution of SHBs
\footnote{the other uncertainties in the expected cumulative 
distribution, see Eq. (3), play a negligible role here},  
which can be constrained at present only through 
the 11 SHBs with known redshift. 

The above estimate compares well with the fraction ($10-30$\%) 
of SHBs produced in GCs, as derived by Grindlay et al. (2006).  
We conclude that the present sample of SHBs with redshift determination 
favors a bimodal origin of SHBs, with high-z bursts resulting 
primarily from the 
merging of primordial NS-NS/BH systems and low-z bursts produced mainly 
by dynamically-formed systems, as first proposed by 
Salvaterra et al. (2008). The contribution  
from the latter systems is unlikely to be $< 10$\%.  
This conclusion might be altered if there existed a large 
population of high redshift  (and thus mainly primordial) SHBs that has so far remained 
unidentified (Berger et al. 2007a, O'Shaughnessy et al. 2008).

On the other hand the association of a SHB with a GC would confirm
unambiguously the dynamical formation scenario. The large offset 
of some SHBs from their host galaxy is consistent with that of GCs\footnote{
%besides GRB050509b and GRB060502b (Gehrels et al. 2005,  Bloom et al. 2006)
%the cases of GRB 061006 and GRB071227 have been recently discussed
%in this perspective (D'Avanzo et al. 2008).},
see Gehrels et al. (2005) for the case of GRB050509b and Bloom et al. (2006) 
for GRB060502b.},
but direct evidence for this is still lacking. This issue may also be addressed
statistically by considering that, even though GCs
are present in galaxies of virtually all types\footnote{except perhaps dwarf
galaxies with $M_V > -13$}, 
most of them are in early type galaxies. Therefore SHBs from dynamically formed 
binaries, besides having on average lower redshifts than those from primordial 
binaries, should be associated preferentially to early type galaxies. 
Since our sample largely overlaps with that of Salvaterra et al. (2008), we 
obtain similar results to theirs in this respect, with 3 in 3 SHBs associated 
to early type galaxies at $z < 0.3$, and 1 in 8 SBHs in association with 
early type galaxies at $z > 0.3$. While these results are in broad 
agreement with a bimodal origin of NS-NS/BH systems giving rise to SHBs,
the low numbers involved prevents us from reaching any firm conclusion. 

The beaming-corrected SHB rate scales as 
\begin{equation}
\rho_0 \simeq 100 (f_b^{-1}/100) (\Delta_1/100)^{\alpha} R_0
\end{equation}
If dynamically formed DNS mergers represent $10$\%
of the total SHB local rate \footnote{Note that the 
contribution from BH-NS mergers is not included.},  
$(f_b^{-1}/100)(\Delta_1/100)^{\alpha} \sim 0.1$
would be required for Eq.~4 to match the approximate merger 
inferred by Phinney et al. (1991) for GC DNSs
($\rho_{DNS}\sim 3 $~Gpc$^{-3}$yr$^{-1}$). 
This indicates that the rough estimate by Phinney et al. (1991) 
is about an order of magnitude lower than
the beaming-corrected rate derived from SHB 
observations, a difference that we do not regard as
crucial at this stage, in consideration of all 
uncertainties.

\section{Prospects for Gravitational Wave Detection}

Within the NS-NS/BH binary interpretation, the local rate of SHBs has clear 
implications for the number of merging events that can be detected with present 
and future ground-based GW interferometers. 
Based on the results in Sect.2, we estimate here 
the number of detectable GW events 
expected in Advanced LIGO/Virgo class interferometers
due to SHBs originating from NS-NS/BH mergers,  

\begin{eqnarray}
{\rm N}_{GW} &\sim & \eta \rho_0[(1-g_{\rm B})V_{\rm DNS}+ g_{\rm B}V_{\rm BHNS})] \\ \nonumber
\simeq &45 \eta & \left(\frac{f_b^{-1}}{100}\right) \left(\frac{R_0}
{4 {\rm Gpc}^{-3}{\rm yr}^{-1}}\right)  \left(\frac{\Delta_1}{100}\right)^{\alpha} \\ \nonumber 
&& [(1-g_{\rm B})+g_{\rm B}\frac{V_{\rm BHNS}}{V_{\rm DNSs}})] {\rm yr}^{-1},
\end{eqnarray}
where $g_{\rm B}$ is the incidence of BH-NS systems among NS-NS/BH progenitors of SBHs,
$V_{\rm DNS}$ and  $V_{\rm BHNS}$ are the volumes corresponding to the horizon of 
GW interferometers for DNSs and BH-NS mergers respectively; the ratio between these 
volumes is $V_{\rm BHNS}/V_{\rm DNS} \sim 10$. 
Here $\eta=1$ for Advanced LIGO/Virgo  and  $\eta=3\times 10^{-4}$  for LIGO/Virgo 
take into account the different volumes sampled by the two instrument classes 
(see e.g. Cutler \& Thorne 2002).
While it is estimated that only $g_{\rm B}\sim 0.01 $ of primordial 
NS-NS/BH systems are BH-NS binaries  
(Belczynski et al. 2007), the fraction of such systems in GCs, though 
presently unknown,   might well be higher, their main formation channel 
being the exchange interaction of a binary containing a NS and a 
cluster star with an isolated BH (Devecchi et al 2007). 
Dynamical star cluster simulations
point to a fairly large population of BH-BH binaries in GCs,
but have not yet provided estimates of $g_{\rm B}$
(O'Leary et al. 2006). For instance for the globular cluster 
parameters adopted by Sadowski et al. (2008) 
NS-NS/BH binaries are expected to form at a very slow pace, 
and it is not surprising that no such systems were produced 
in those simulations.

We note that as long as $g_{\rm B}\gsim 0.15 $ the rate 
of detectable GW events from BH-NS binaries will be higher 
than that from DNSs (see Eq. 5). 
We assume $g_{\rm B} = 0.5 $ 
(and the other fiducial values in Eq.(5))
in the estimates of $N_{GW}$ given below, and report also 
in parentheses the values corresponding to $g_{\rm B}\sim 0$ 
and $1$, respectively. 
For Advanced LIGO/Virgo we find that events from primordial 
and dynamically formed NS-NS/BH binaries are expected  
at a rate of $\sim$ 14~yr$^{-1}$ 
and $\sim 248$~yr$^{-1}$ (45, 450), respectively. 
For $g_{\rm B}= 0.5 $ the latter event rate is dominated by merging 
BH-NS binaries formed in GCs ($\sim 226$~yr$^{-1}$) and is close to the 
upper end of the range 
estimated by Kalogera et al. (2006) for primordial DNSs.
For present generation interferometers Eq.~5 gives 1 event in 238 years  and 
1 event in 13 years (74, 7 years) from primordial and dynamically 
formed systems, respectively.

There are presently substantial uncertainties 
in the values of the parameters in Eq.(5) and therefore 
a precise estimate of the expected number of detectable GW event cannot be made yet. 
However it should be possible in the near future to constrain more tightly the
uncertain parameters in Eq (5). Concerning the beaming factor,  
a few additional detailed studies of the optical afterglow of SHBs
with currently available instrumentation will allow a more precise 
determination of $f_{b}^{-1}$.  Taken at face value, present 
estimates range over a factor of $\sim 2.6$. 
The local SHB event rate $R_0$, while 
constrained by the large sample of SHBs revealed by BATSE ,
can vary up to a factor of $\sim 1.8$ depending on 
the redshift distribution and incidence of dynamically formed SHBs 
(see Sect. 2.1 and 2.2).  The latter 
is presently determined through 11 SHBs, but the 
sample of SHBs with secure redshift is steadily increasing 
and should triple by the end of the Swift mission. 
A more accurate determination of the lower end of the 
SHB luminosity function will likely require more sensitive
GRB detectors than currently available; however, as we discussed in
Sect. 2.1 the lower limit $L_{min}$ will likely decrease  
({\it i.e.} $\Delta_1$ will increase) resulting in a higher  
expected rate of detectable GW events. 
The incidence of NS-BH systems among dynamically 
formed SHB progenitors $g_{\rm B}$ can be determined through  
more extensive dynamical GC simulations.  The whole 
range of allowed values of $g_{\rm B}$ ({\it i.e.} $0 \div 1$) 
translates into a factor of $\sim 10$ uncertainty in the 
number of GW events in Eq.(5). 

Our treatment assumes that SHBs with similar properties 
are produced both by DNS and BH-NS mergers. From the point of view of the models,
this appears to be a reasonable ansatz as the torus of matter at nuclear densities is 
expected to originate from the lightest collapsed object in the system 
and provide the main source of energy for the SHB. 
Moreover, observations have not yet shown clear evidence 
that SHBs comprise different subclasses. Alternatively, if all SHBs 
detected so far came from DNSs, the total GW event rate from 
dynamically formed DNS and BH-NS mergers would be higher 
(unless $g_B=1$ holds).  On the more pessimistic side, 
if a fraction $h$ of SHBs were due to an entirely different
phenomenon ({\it e.g.} giant flares from Soft Gamma Repeaters, Hurley et al. 2005) 
all the rates given above would be decreased by a factor $\sim (1-h)$. 

According to Sadowski et al. (2008), the rate of 
BH-BH mergers formed in globular clusters is high and can give rise to 
0.01-1 event per year in LIGO and 25-3000 events per year in Advanced LIGO,
a major uncertainty being the initial stellar 
mass fraction in dense clusters. Other authors envisage substantially
lower rates (Portegies Zwart and McMillan 2000, O'Leary et al. 2006).
Once the ratio between NS-NS/BH and BH-BH 
binaries in globular clusters will be better known from simulations, 
the merging rate of dynamically formed NS-NS/BH binaries inferred from SHBs 
can provide an independent normalization for the BH-BH merger rate.

We note that an interferometer's horizon would increase by a factor of $\sim 2.4$  
for a GW detection in coincidence with a SHB detection (Cutler \& Thorne 2002). 
This is because of the higher sensitivity afforded by having an independent 
knowledge of the time of occurrence and sky position of the event. 
The above factor takes also into account that, if SHBs are beamed 
along the binary's angular momentum axis, then the SHBs that are detected 
at the earth are also the ones whose orientation angle maximises the GW 
signal. The larger horizon translates into a factor of $\sim 15$ increase in 
both $V_{\rm DNS}$ 
and $V_{\rm BHNS}$. Considering that only 
a fraction $f_b\sim 0.01$ of the SHBs are beamed toward us, the coincident event rate  
would be $\sim 0.15 (f_b^{-1}/100)$ of the rates derived from Eq.~5. 

\section{Conclusions}

In this paper we have presented a simple method for inferring the merging 
rate and corresponding rate of detectable GW signals from coalescing 
NS-NS/BH binaries. The method is based on the assumption that at least a 
sizable fraction of SHBs originates in the merging of such binaries and
exploits the most recent results from observations to infer the luminosity function, 
redshift distribution and degree of beaming of SHBs. We find further evidence in 
favor of a bimodal origin of SHB progenitors, with the merging of primordial 
binaries dominating at high redshift, whereas at lower redshift at least 
$\sim 10$\%  (and probably about half) of the events arises from coalescing binaries formed 
dynamically in GCs. We have shown that the latter events make the 
expected local merging rate higher.

The accuracy of the GW event rate that can be estimated at present is 
hampered by our present knowledge of the parameters in Eq.(5). 
 Maximum uncertainties in the SHB beaming factor ($f_b$) and local rate ($R_0$) 
add up to a factor of $\sim 5$, whereas the ignorance of the incidence
$g_B$ of BH-NS binaries among dynamically formed systems adds 
a factor of $10$ uncertainty. This results in total uncertainty 
by a factor of $\sim 50$ in the detectable GW event rate in Eq.~(5). 
Though large, this is not distant from the uncertainties of 
other estimates of GW events in literature.  
For instance Kalogera et al. (2004) estimate the galactic double neutron 
star merger rate with an uncertainty of factor of ~14.  
Similarly, the results of the dynamical simulations of Sadowski et al. 
(2008) for BH-BH mergers are affected by a ~1.5 decade wide uncertainty 
in the initial mass fraction of globular clusters. 
We expect that further SHBs observations in Swift era will lead to a more 
accurate determination of $f_b$ and $R_0$, while more advanced dynamical globular 
cluster simulations will decrease the range of allowed values of $g_B$.  
Progress in this area might well take place at a sufficiently fast pace,  
that our method can yield accurate predictions before GW events
from coalescing binaries are detected in large numbers. 

We have shown that a sizeable fraction of 
detectable GW events is expected to be coincident with SHBs
(depending on the SHB beaming factor): this provides a new 
interesting perspective for the Advanced LIGO/Virgo era. 
We conclude that SHBs can provide a powerful means of inferring the GW event 
rate from coalescing binaries containing two collapsed objects.

\acknowledgements
We acknowledge useful discussions with Monica Colpi, Livia Origlia and Andrea Possenti. 
This work was partially supported through ASI/INAF
contracts ASI/I/R/039/04 and ASI/I/R/023/05.

%\begin{references}

\begin{table}[t]
\caption{Model parameters and local rates}
\begin{tabular}{lllll}
 \hline
DNSs (BH-NS) & $L^*$ & $\alpha$ & $\beta$ & $R_0$ Rate(z=0)\\
   & $[10^{51}\erg\pers]$   &  & &  Gpc$^{-3}$yr$^{-1}$\\
  \hline
Primordial          & $2 $     & 0.6   & 2      & 1.3 \\
Dynamically formed  & $0.8$    & 0.8   & 2       & 4.0\\
 \hline
\label{t:fit}
\end{tabular}
\end{table}

%\begin{figure}
%\includegraphics[width=1.0\columnwidth, keepaspectratio]{f1.eps}
%\figcaption[FileName]{\label{cdf} Observed (histogram) and expected
%(curves) cumulative $z$-distribution functions for the 
%different time-delay probabilities expected in the two scenarios.
%Model distributions are shown for different 
%combination of primordial and dynamically
%formed NS-NS/BH systems.}
%\end{figure}

\begin{figure}
%\centering \noindent
{\par\centering \resizebox*{0.95\columnwidth}{!}
{\includegraphics
%[width=8cm,height=8cm]
{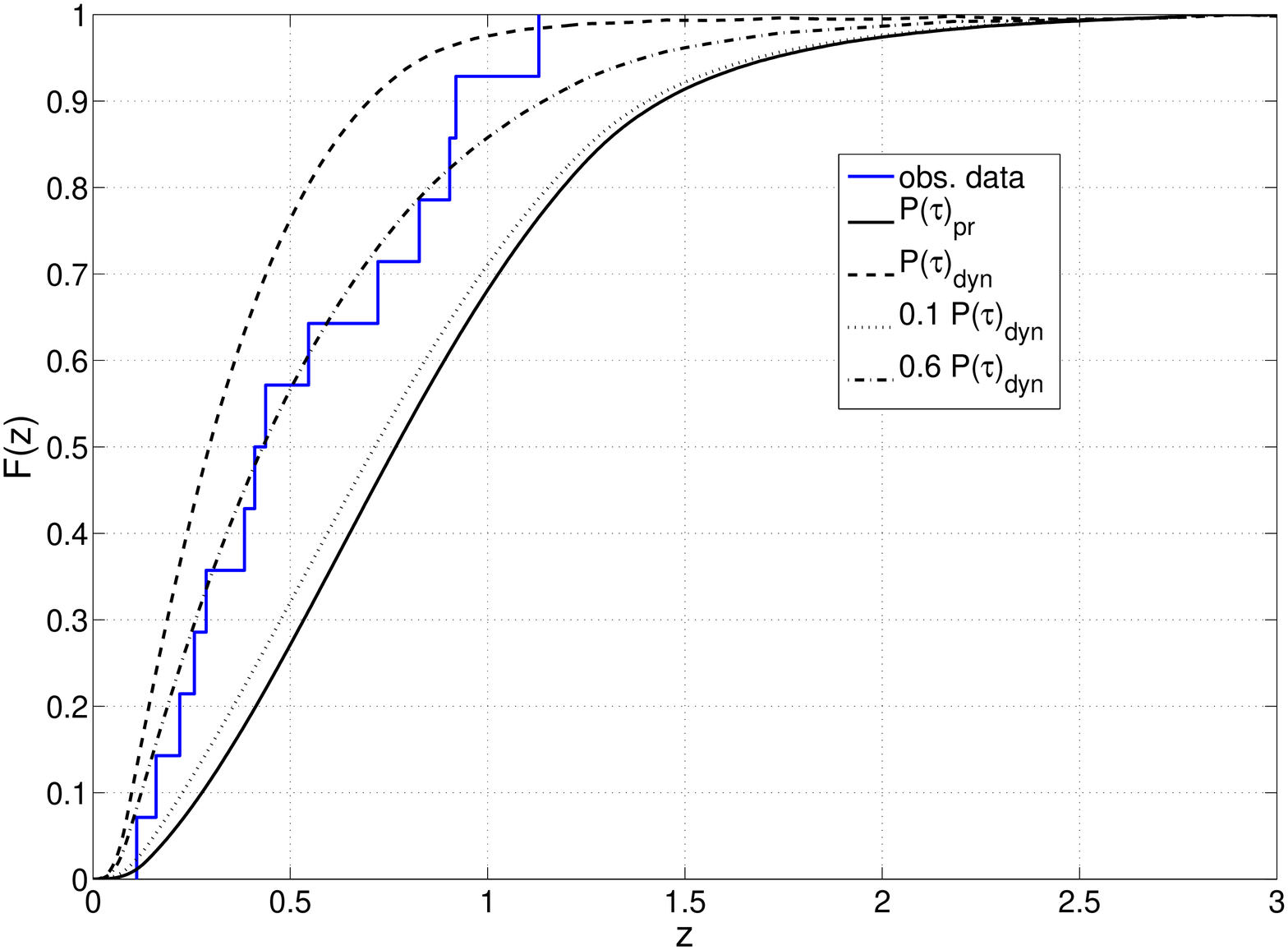}} \par} \caption{\label{cdf} Observed (histogram) and expected
(curves) cumulative $z$-distribution functions for the 
different time-delay probabilities expected in the two scenarios.
Model distributions are shown for different 
combination of primordial and dynamically
formed NS-NS/BH systems. } \label{cdf}
\end{figure}

\end{document}